\documentclass[twocolumn,prl,floatfix,superscriptaddress]{revtex4}
\usepackage{graphicx}
\usepackage{amsmath}
\usepackage{latexsym}
\usepackage{epstopdf}
\usepackage{amsmath}
\usepackage{amssymb,amsmath}

\newcommand{\beq}{\begin{equation}}
\newcommand{\eeq}{\end{equation}}

\begin{document}

\title{Discrete scale invariance and stochastic Loewner evolution}
\author{M. Ghasemi Nezhadhaghighi$^*$}
\affiliation{Department of Physics, Sharif University of Technology, Tehran, P.O.Box: 11365-9161, Iran}
\author{M.~A.~Rajabpour}
\affiliation{SISSA and INFN, \textit{Sezione di Trieste},  via Bonomea 265, 34136 Trieste, Italy}

\begin{abstract}
In complex systems with fractal properties the scale invariance has an important rule to classify different statistical properties. In two dimensions the Loewner equation can classify all the fractal curves. Using the Weierstrass-Mandelbrot (WM) function as the drift of the Loewner equation we introduce a large class of fractal curves with discrete scale invariance (DSI).
We show that  the fractal dimension of the curves can be extracted from the diffusion coefficient of the trend of the variance of the WM function. We argue  that, up to the fractal dimension calculations, all the WM functions follow the behavior of the corresponding Brownian motion. Our study opens a way to classify all the fractal curves with DSI.
 In particular, we investigate the contour lines
of $2D$ WM function as a physical candidate for our new stochastic curves.

\end{abstract}

\maketitle
\section{Introduction}Discrete scale invariance (DSI) is a weaker kind of scale invariance
according to which the system obeys scale invariance $\mathcal{O}(\lambda x)=\mu(\lambda) \mathcal{O}(x)$ only for specific choices of $\lambda$.
The list of applications of discrete scale invariance (DSI) covers phenomena in  programming and number theory,
diffusion in anisotropic quenched random lattices, growth processes and rupture, quenched disordered systems, turbulence, cosmic lacunarity, etc,
for review of the concept and applications see \cite{sornette1,sornette2}. In some cases the DSI is manifested in the geometry of the system, the most
famous cases being animals \cite{SA} and Diffusion-limited aggregation (DLA) \cite{WS} . The fractal dimension of these models shows log-periodic corrections to scaling which is
the signature of the presence of complex exponents and DSI \cite{SS,SJMS}.\newline One of the methods to tackle these models, especially the DLA, is the iterative conformal mapping technique introduced by Hastings and Levitov in \cite{HL}. It is based on approximating an addition of a particle to preexisting domain with bumps at nearby positions produced by conformal maps. This method was used successfully to describe the Laplacian stochastic growth models \cite{HL} as well as the non-Laplacian transport processes \cite{BCD}, for a recent review see \cite{BC}.\newline
Another method, in the same spirit as the Hastings and Levitov method, to study geometrical critical systems is by Schramm-Loewner evolution \cite{Schramm}, for  review see \cite{bernard0}.
 It is a family of random planar curves with conformal symmetry that can be described by successive random conformal maps.
This method recently has found lots of applications in the physical
systems such as domain walls in statistical models \cite{bernard0}, zero-vorticity lines of Navier-Stokes turbulence \cite{bernard} and
 iso-height lines in rough surfaces \cite{rajab,Schramm Sheffield}. \newline Although SLE gives a powerful way to classify all the conformally invariant curves in two dimensions it is still too restrictive to classify the curves with weaker symmetries such as scale invariance. \begin{figure} [htb]
\centering
\includegraphics[width=0.5\textwidth]{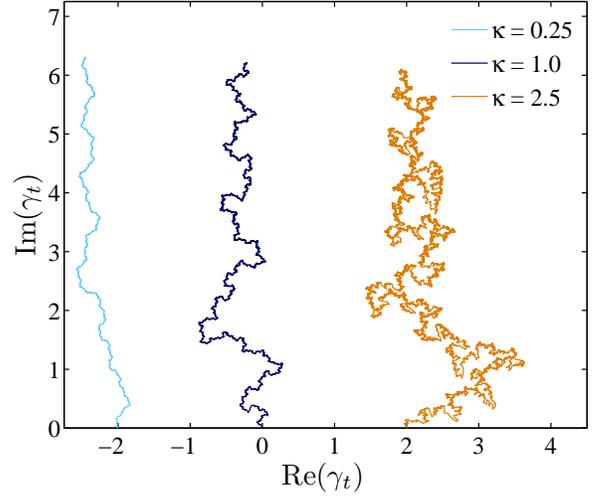}
%\vspace{5cm}  % amount of vertical space needed
\caption{(Color on line) Curves, from left to write $\kappa=0.25,1,2.5$, generated by taking WM function with $\lambda=2.5$.
The total time is $T=10$ and the number of points is $N=3\times 10^4$. The computer running time for every realization is ~15 min.}
\label{Figure:1}
\end{figure}In this letter we will introduce a broader range of fractal curves in two dimensions.  The method takes the advantage of the Brownian motion limit of the WM function and opens a way to classify the DSI fractal curves in two dimensions.\newline
The idea behind SLE formulation is parameterizing a growing curve with time $t$
in a two dimensional domain with some successive conformal maps that
can remove the curve. The most familiar case is the  curve growing
with the tip at $\gamma_{t}$ in the upper half plane $\mathbb{H}$.
The curve can be parametrized with the conformal map
$g_{t}(z):\mathbb{H} \setminus K_{t}\rightarrow \mathbb{H}$ which
maps the upper half plane minus the hull \cite{footnote} $K_{t}$,
 to the upper half plane
and obeys the following Loewner equation
\begin{eqnarray}\label{Loewner}
dg_{t}(z)=\frac{2dt}{g_{t}(z)-U(t)},
\end{eqnarray}
where $U_{t}$ is the drift of the process and can be an arbitrary
continuous function with H\"older exponent $h\geq \frac{1}{2}$.
\begin{figure} [htb]
\centering
\includegraphics[width=0.44\textwidth]{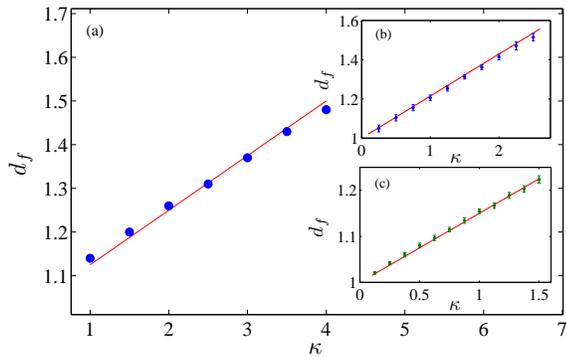}
%\vspace{5cm}  % amount of vertical space needed
\caption{ (Color on line) The fractal dimension of the generated curves versus
$\kappa$ for different drifts: $a$) ordinary SLE with Brownian
motion as a drift, $b$) $WSLE(e^{ix},2.5,\kappa)$ and $c$)
$WSLE(sin^{-1}(sin(x)),2.5,\kappa)$. The continuous lines are the
lines coming from the equation (\ref{kappaeffective}). In $a$ the
number of realizations for every $\kappa$ is ~1000 and the error
bares are smaller than the diameter of the circles. In $b$ and $c$
the number of realizations for every $\kappa$ is ~500. The number of
points on every realization was $N=3 \times 10^4$.} \label{Figure:2}
\end{figure}
Every growing curve is related to a special $U(t)$ and by having
$U(t)$ one can extract a growing curve in the upper half plane.
Considering $U_{t}$ as a process proportional to the Brownian motion
$\sqrt{\kappa}B_{t}$ one can extract conformal invariant curves and
the evolution is called SLE \cite{Schramm}. Brownian motion is a
Gaussian Markov stochastic process with mean zero and
$E[B_{t}B_{s}]=min(t,s)$. All conformal invariant curves in two
dimensions can be parametrized by a Brownian motion. \newline The
fractal dimensions of the curves are linearly proportional to
$\kappa$ and given by \cite{bernard0,Beffara}
\begin{eqnarray}\label{fractal dimension}
d_{SLE}=1+\frac{\kappa}{8}.
\end{eqnarray}
The linearity of the relation between fractal dimension and $\kappa$
comes from the specific form of the Fokker-Planck equation of the
process which is a second order differential equation.
The natural question is what is the appropriate $U(t)$ for the
curves with DSI? To answer this question consider a drift with DSI
as $\tilde{U}(t):=\frac{1}{\lambda}U(\lambda^{2}t)$, where
$\lambda$ is a positive real number,  then one can show that
$\tilde{g}_{t}(z)=\frac{1}{\lambda}g_{\lambda^{2}t}(\lambda z)$
satisfies the same Loewner equation as (\ref{Loewner}). To see this consider 
the new time parametrization in equation (\ref{Loewner}) as $t \to a(t)$. Then one can write the equation (\ref{Loewner}) for the general drift as $d\hat{g}_{t}(z)=\frac{2\frac{da(t)}{dt}}{\hat{g}_{t}(z)-U(a(t))}$, where $\hat{g}_{t}(z)=g_{a(t)}(z)$ and $\hat{g}_{0}(z)=z$. Considering $a(t)=\lambda^{2}t$ it is easy to get
\begin{eqnarray}\label{Loewner2}
d\tilde{g}_{t}(z)=\frac{2dt}{\tilde{g}_{t}(z)-\frac{1}{\lambda}U(\lambda^{2}t)},
\end{eqnarray}
where $\tilde{g}_{t}(z)=\frac{\hat{g}_{t}(z)}{\lambda}$ and $\tilde{g}_{0}(z)=z$. Since $\tilde{U}(t)$ is the same as $U(t)$, in the distribution sense, for our DSI process one can conclude that $\tilde{g}_{t}(z)$
satisfies the same Loewner equation as (\ref{Loewner}).
\begin{figure} [htb]
\centering
\includegraphics[width=0.45\textwidth]{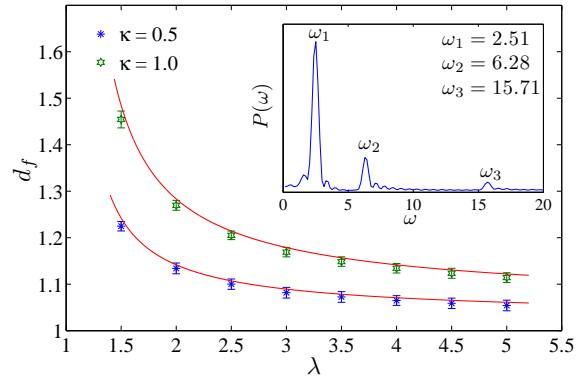}
%\vspace{5cm}  % amount of vertical space needed
\caption{ (Color on line) The fractal dimension of the generated curves versus
$\lambda$ for two different $\kappa$s for $WSLE(e^{ix},\lambda,1)$(top) and $WSLE(e^{ix},\lambda,0.5)$ (bottom). The number of realizations for every
$\lambda$ is 50. Inset: The Lomb periodogram of the Re$(\gamma_t)$, here $\omega_{n}$ is equal to $\lambda^n$.}
\label{Figure:3}
\end{figure}
Using this
property one can claim that  $\tilde{g}_{t}(z)$ is the conformal map
that takes the complement of the hull
$\tilde{K}_{t}:=K_{\lambda^{2}t}/\lambda$ and maps it to the upper
half plane. The natural consequence of taking discrete scale
invariant drift in the Loewner equation is getting a discrete scale
invariant curve in the upper half plane and vise versa. With this
introduction, in this paper we will take the  generic
process with  DSI, Weierstrass-Mandelbrot (WM) type functions
\cite{BL,Falconer, Sornette3}, as a drift and we will investigate numerically
the fractal properties of the corresponding curves in the upper half
plane. We will also introduce the contour lines of the $2D$ WM
functions as the new fractal curves with DSI and as the possible
candidates for our new stochastic Loewner evolution.

\section{Loewner equation with DSI drift}
The building block of the functions with DSI is Weierstrass-Mandelbrot (WM) function
which was studied by Berry and Lewis in \cite{BL}. It is a random continuous non differentiable
mono fractal function that one can see it in a wide range of physical phenomena such as
sediment and turbulence \cite{sornette1,sornette2}, fractal properties of the landscapes and other enviromental data \cite{Burrough}, contact analysis of elastic-plastic fractal surfaces \cite{Yan} and propagation and localization of waves in fractal media \cite{Localiz}. The WM function can be defined as follows
\begin{eqnarray}\label{WM}
W(h,t)=\sum_{n=n_{min}}^{\infty}\lambda^{-nH}(h(0)-h(\lambda^{n}t))e^{i \phi_{n}},
\end{eqnarray}
 where $h(t)$ can be any periodic function which is differentiable at zero and $\phi_{n}$ is an arbitrary phase chosen from the interval $[0,2\pi]$, to make the series convergent we need
to consider $0<H<1$. It was proved in \cite{BL,Taqqu} that the above process with $h(x)=e^{ix}$
converges to the fractional Brownian motion in the limit $\lambda\rightarrow 1$ . In our numerical
calculation we will take $h(x)=e^{ix}, \sin^{-1}(\sin(x))$, however, our results are extendable to the most generic cases.
The WM process is not Markov but has stationary increments. It is a fractal function for the finite $n_{min}$ and  also for $n_{min}=-\infty$, however,
 it has full DSI just for
$n_{min}=-\infty$. The fractal dimension of the function is $2-H$
where $H$ is the H\"older exponent. 

 Taking Re$
(\sqrt{\kappa}W(h,t))$ as the drift of the Loewner equation one can get simple curves just
for $H\geq \frac{1}{2}$. The fractal dimension of the curves is ~1
for $H>\frac{1}{2}$ but non-trivial for $H=\frac{1}{2}$. From this
moment, we will consider just $H=\frac{1}{2}$ which has the same
fractal dimension as the Brownian motion but it is not a Markov
process. One can extract the corresponding curves numerically as an
iteration process of infinitesimal conformal mappings as follows: $
\gamma(j\tau)=f_{1}\circ f_{2}\circ...\circ f_{j}(\xi_{j})$, where
$\xi_{j}$ is the drift, in our case WM process, and
$f_{j}(z)=\sqrt{(z-\xi_{j})^{2}-4\tau}+\xi_{j}$ is the inverse
Loewner map that can produce a small slit at $\xi_{j}$. 
The discretized driving force is constant in the interval
$[(j-1)\tau,j\tau]$ where $\tau=\frac{t}{N}$ is the mesh of the time
and $N$ is the number of points on the curve. We will call these new
curves $WSLE(h,\lambda,\kappa)$. In Fig~1 one can see
$WSLE(e^{ix},2.5,\kappa)$ for three different $\kappa$'s. 
In Fig ~2
the box fractal dimension of SLE,  $WSLE(e^{ix},2.5,\kappa)$ and
$WSLE(sin^{-1}(sin(x)),2.5,\kappa)$ were shown with respect to
$\kappa$ and we observe that they are all linearly dependent on
$\kappa$.
The slopes of the curves in Fig~2
is dependent on the function $h(x)$ and $\lambda$. 

 One can  find a theoretical  approximation for the fractal dimension of the extracted curves by calculating
the second moment of the real part of the WM function. 
Using Poisson re-summation formula one can find the following formula
for the trend of the WM process
\begin{eqnarray}\label{trend}
|\textrm{Re}(\sqrt{\kappa}W(h,t+\tau))-\textrm{Re}(\sqrt{\kappa}W(h,t))|^{2} \approx \nonumber\\
\frac{\kappa\tau}{2\ln \lambda}\int_{0}^{\infty}|\frac{h(0)-h(x)}{x}|^{2}dx.
\end{eqnarray}
\begin{figure} [htb]
\includegraphics[width=0.5\textwidth]{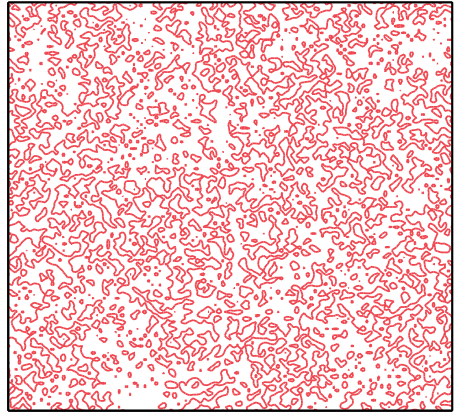}

%\vspace{5cm}  % amount of vertical space needed
\caption{(Color on line) The contour lines of the $2D$ WM function with $\lambda=1.5$, $M=20$ and $L=200$.}
\label{Figure:4}
\end{figure}
The above equation shows that $\sqrt{\ln\lambda }$ Re$(W(h,t))$
converges to the Brownian motion in the limit $\lambda\rightarrow1$.
Then if we consider the WM process as a non-Markovian perturbation of
the Brownian motion, one can introduce an effective diffusion
constant
\begin{eqnarray}\label{kappaeffective}
\kappa_{eff}=\kappa \frac{1}{2\ln \lambda}\int_{0}^{\infty}|\frac{h(0)-h(x)}{x}|^{2}dx,
\end{eqnarray}
 Assuming the universality of the coefficient $\frac{1}{8}$ the fractal dimension will be the same as (\ref{fractal dimension})
but with the $\kappa$ substituted with $\kappa_{eff}$. It turns out
that the above approximation gives convincing results, in Fig~2 the
continuous lines are the lines coming from this approximation. To
check the
 relation between fractal dimension and $\lambda$ we calculated the fractal dimension for $\kappa=0.5$ and $1$ for the different $\lambda$s,
the results are shown in Fig~3. Up to the numerical accuracy the
proposed formula is perfectly confirmed. This result shows that as long as the fractal dimension calculations is concerned, the Brownian motion behaves
like a fixed point of the all the DSI processes with  $H=\frac{1}{2}$.
In the same figure we put also the result of the Lomb test
\cite{lomb} on the Re$(\gamma_t)$ which shows the DSI properties of
the curves.

In the rest of the paper we will introduce some new DSI curves that
can be studied with $WSLE(h,\lambda,\kappa)$. 

\section{Contour lines of the DSI surfaces}

It was shown in
\cite{Schramm Sheffield} that the contour lines of Gaussian free
field theory (GFF) can be described by $SLE(\kappa,\rho)$. In other words
the contour lines of rough surfaces with the zero roughness exponent
are conformally invariant and can be described by the Schramm's
method. Having this fact as a motivation we generated a $2D$ WM
function \cite{AB}
\begin{eqnarray}\label{2DWM}
z(x,y)=(\frac{\ln \lambda}{M})^{\frac{1}{2}}\sum_{m=1}^{M}A_{m}\sum_{n=-\infty}^{\infty}(\frac{2\pi \lambda^{n}}{L})^{-H}
\Big(\cos\phi_{m,n}-\nonumber\\
\cos\big(\tfrac{2\pi \rho \lambda^{n}}{L}\cos(\theta-\alpha_{m})+\phi_{m,n}\big)\Big),
\end{eqnarray}
\begin{figure} [htb]
\centering
\includegraphics[width=0.5\textwidth]{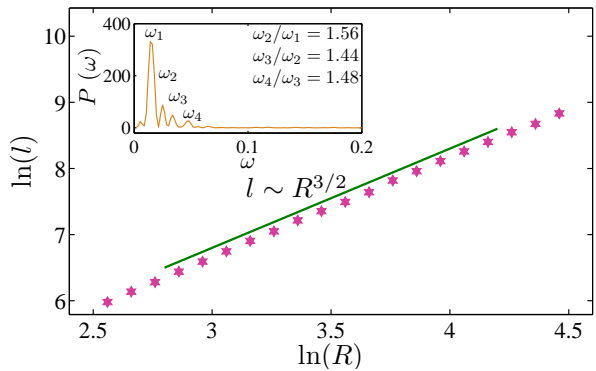}
%\vspace{5cm}  % amount of vertical space needed
\caption{ (Color on line) The perimeter of the contour lines $l$ with respect to the
radius of gyration $R$. To get the graph we averaged over all the
perimeters of the contour lines with the radius of gyration in the
interval $[\textrm{ln}(R),\textrm{ln}(R)+\delta]$.  Inset: The Lomb periodogram of
the contour lines.} \label{Figure:5}
\end{figure}
where $\rho=\left(x^2+y^2\right)^{1/2}$ and $\theta=tan^{-1}\left(\frac{y}{x}\right)$. The phase $\alpha_{m}=\pi m/M$ and $A_{m}$ make the system homogeneous and $\phi_{m,n}$ , which is randomly chosen from the
interval $[0,2\pi]$, makes the system random. For the isotropic surfaces we have
$A_{m}=const$. In numerical methods the two parameters $M$ and $n$ and the size of the system $L$ should be finite but large enough that $z\left(x,y\right)$  does not change with increasing these parameters. The roughness exponent and the fractal dimension $D=3-H$ are two important parameters in simulating WM function. The surfaces with larger $H $ seems smoother than the surfaces with smaller $H $. The $2D$ WM
function obeys the scaling law $z(\lambda x,\lambda y)=\lambda^{H}z(x,y)$. This shows that $z(x,y)$ is a self similar function with respect to the scale $\lambda$.

We generated $750$
realizations of the above surfaces with $H=0$, $L=1024$, $M=20$ and
$\lambda=1.5$ then we constructed the contour lines of the surfaces
(as already explained in \cite{KH, RV}) at the average height, see
Fig.~4. 
The average fractal dimension of the roughly $10^6$ contour
lines is $D=1.505\pm0.005$, see Fig ~5, which is very close to
$\frac{3}{2}$. In the next step we took the largest contour lines
that do not cross the boundary and found
$|\tilde{\textbf{r}}_{i}|$, where
 $|\tilde{\textbf{r}}_{i}|=|\textbf{r}_{i}-\textbf{R}|$ is the distance of the points of the contour $r_{i}$ from the center of mass $\textbf{R}$.
This parameter shows very clear DSI after using Lomb periodogram, Fig ~5.\newline
To check the DSI of the drift of the Loewner equation directly, we
first consider an arbitrary placed horizontal line representing the
real axis in the complex plane across the $2D$ WM profiles. Then we
cut the portion of each curve $\gamma_{t}$ above the real line as it
is in the upper half plane $\mathbb{H}$. 
There are different numerical
algorithms to extract the driving function of a given random curve
by inversion of the Loewner equation \cite{TK}. \begin{figure} [htb]
\centering
\includegraphics[angle=00,width=0.45\textwidth]{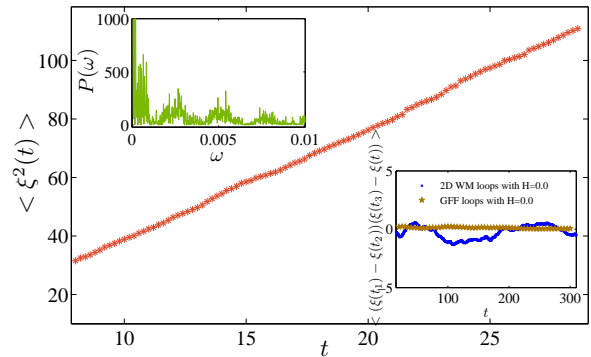}
%\vspace{5cm}  % amount of vertical space needed
\caption{ (Color on line) The variance of the drift of the Loewner equation for the
contour lines of $2D$ WM function. Top inset: the Lomb
periodogram of the drift. Bottom inset: the correlation of the increments of the drift of the contour lines of the GFF and the contour lines of the $2D$ WM function. The number of realizations were approximately $7000$ and the number of points on every realization was $3000-4000$. } \label{Figure:6}
\end{figure}
We have used the
successive discrete, conformal slit maps – based on the piecewise
constant approximation of the driving function – that swallow one
segment of the curve at each time step \cite{Nodalline}. To do this,
at first, we have determined the contour lines  of the $2D$ WM
surfaces as described above by the sequences of points ${z_{0},
z_{1},..., z_{N}}$. We obtained ~2000 such curves with an average
number of points about ~5000. After that all of the contour lines
were mapped by $\varphi(z) = z_{N}z/(z_{N}-z)$. To avoid numerical
errors only the part of the curves corresponding to capacity $8-300$
were used. It is worth to mention here that since we do not expect
to have conformal invariance for our contour lines the effect of the
map $\varphi(z)$ on the measure of the curves is not clear. Since
this map is a global conformal map we do not expect a huge
difference between the measure of curves before and after the
mapping. Fig.~6 shows the properties of the drift calculated with
the above method. The drift shows clear DSI in the Lomb periodogram with $\lambda=1.47\pm0.09$,
moreover the variance of the drift is linear around the line with
slope $\kappa=3.9\pm 0.2$ which, up to the numerical errors, is very
close to ~4, see Fig ~6. The increments of the extracted drift shows oscillating non-zero correlation which is clearly indicating that the drift is not Markovian, in  Fig ~6 we compared these results with those coming from the simulation of the contour lines of the Gaussian free field theory. This comparison shows that the contour lines of $2D$ Weierstrass
function with $H=0$ could be related to the Loewner
equation with the WM function as a drift. Here it is worth mentioning that we expect the same results for the $2D$ WM functions generated by the other periodic functions insteed of $\cos(x)$, we will discuss  properties of the contour lines of 2D WM function for generic $H$ for the different periodic functions in the forthcoming paper \cite{GR}. 

\section{Conclusion} We made the first attempt to classify and study systematically the fractal curves with DSI based on Loewner evolution. Our approach is from many sense reminiscent of the renormalization group and universality ideas in statistical mechanics and field theory. By looking to the Brownian motion as the fixed point of all the WM functions with $H=\frac{1}{2}$ we showed that the fractal dimension of the fractal curves follows the behavior of the fixed point. Our approach supports the idea of looking to the scale invariant curves as the perturbation of the conformally invariant curves. We also
extracted contour lines of $2D$ fractal surfaces  as  physical
examples for the fractal curves with DSI. In
addition the corresponding drift of the Loewner equation for these
curves was extracted by using numerical methods. The drift shares
many properties with the WM process. \newline It is interesting to understand the connection of our approach to the older studies bases on iterative conformal mappings \cite{HL}, see also \cite{SL}. The DSI is manifested mostly because of the hierarchical underlying graphs, our study shows that it may be possible to study statistical models, on graphs with the well-defined continuum limit, which show DSI at the critical point. Investigating statistical models with such a property could be interesting.  

\begin{center}
{\bf ACKNOWLEDGMENTS}

We thank F. Franchini and M. Kardar for reading the manuscript and D. Sornette for taking our attention to the reference \cite{Sornette3}. We also appreciate providing computing facilities by MPI cluster of SUT. M. A. Rajabpour thanks S. Sheffield for discussion. M. G. Nezhadhaghighi thanks S. Sheykhani for computer support.

\end{center}


\begin{thebibliography}{99}
\bibitem{footnote}By hull we mean
the curve and the region of the upper half plane which is separated
from infinity by the curve
\end{thebibliography}
\end{document}